\documentclass{PoS}
\usepackage{graphicx}
\usepackage{myaasmacros}

\title{Quest for detection of a cosmological signal from neutral hydrogen with a digital radio array developed for air-shower measurements}

\ShortTitle{Tunka-21cm project}

\author{
\speaker{D.~Kostunin}$^{1}$, 
P.~Bezyazeekov$^{2}$, 
N.~Budnev$^{2}$, 
O.~Grishin$^{2}$, 
O.~Fedorov$^{2}$, 
Y.~Kazarina$^{2}$, 
L.~Kuzmichev$^{3}$, 
S.~Malakhov$^{2}$, 
T.~Marshalkina$^{2}$, 
V.~Oreshko$^{4}$, 
M.~Pshirkov$^{4,5,6}$, 
G.~Rubtsov$^{6}$, 
A.~Sokolov$^{6,7}$,
A.~Zagorodnikov$^{2}$,
D.~Zhurov$^{2}$
~\\
$^{1}$DESY, Platanenallee 6, 15738 Zeuthen, Germany\\
$^{2}$Applied Physics Institute ISU, 664020 Irkutsk, Russia\\
$^{3}$Skobeltsyn Institute of Nuclear Physics MSU, 119991 Moscow, Russia\\
$^{4}$P.N. Lebedev Physical Institute, Pushchino Radio Astronomy Observatory, 142290 Pushchino, Russia\\
$^{5}$Sternberg Astronomical Institute, Lomonosov Moscow State University, 119992 Moscow, Russia\\
$^{6}$Institute for Nuclear Research RAS, 117312 Moscow, Russia\\
$^{7}$Institute of Theoretical and Experimental Physics, 117218 Moscow, Russia
~\\
E-mail: \email{tunka-21cm@astroparticle.online}
}
\abstract{Digital radio arrays are widely used for the low-frequency radio astronomy as well as for detection of air-showers induced by high-energy cosmic rays and neutrinos.
	Since the radio emission from air-showers forms short broadband pulses with duration of tens nanoseconds, the data acquisition strategies of cosmic-ray and astronomical arrays have significant differences.
	To perform precise measurement of cosmic rays, the radio array should have absolute amplitude calibration and record the entire electric field on the antenna in the broad frequency range.
	These requirements are similar to ones defined for the experiments aimed at the detection of weak signal from neutral hydrogen  at redshifts of $z$>10, what led us to the application of our experience with Tunka-Rex to this problem.
	We are developing new experimental setup comprising of four antenna stations, placed on the area of 100 sq.m.
	Each antenna station consists of two perpendicular loop antennas measuring electric field in the frequency band of 30-80 MHz.
	The setup records electric fields from all antennas in portions of 50 $\mu$s reaching the spectral resolution of 20~kHz.
	We expect a flow of redundant data of about 10 GB/day, and plan to exploit this redundancy in order to decrease systematic uncertainty of the measurements by application of digital beam-forming, matched filtering and RFI suppression with neural networks.
	In the present contribution we describe the design and calibration of the setup, expected performance and data analysis techniques.}

\FullConference{36th International Cosmic Ray Conference -ICRC2019-\\
		July 24th - August 1st, 2019\\
		Madison, WI, U.S.A.}
\begin{document}

\section{Introduction}
The cosmological signal coming from the line of neutral hydrogen could be  a precious source of valuable information from early stages of the evolution of the Universe  from the Epoch of recombination ($z\sim1100$) to the end of the Epoch of Reionization (EoR, $z\sim6$). Due to the complex interplay of different astrophysical processes there is still no complete theoretical description of the signal and its evolution with redshift. 
If at some redshift $z$ there is a difference between spin temperature of the hyperfine structure of neutral hydrogen, and  temperature of the radiation source -- cosmic background, certain spectral feature would arise, that can be potentially observed at frequency $\nu=1420/(1+z)~\mathrm{MHz}$ \cite{Furlanetto2006,Morales2010}. The spin temperature describes the relative population of different hyperfine levels and can be set by either radiation temperature of the temperature of the surrounding medium. These two temperatures could differ which in turn would lead to emergence of the signal.

More specifically, transition between levels and thus spin temperature can couple to the temperature of the medium via collisions between atoms or to the background photon field in the Ly-${\alpha}$ line (Wouthuysen-Field effect), the latter begins to operate effectively after formation of first stars after $z\sim20$.  During earlier epochs gas temperature due to the  adiabatic expansion was lower than the temperature of the background radiation and that results in a potentially observable shallow drop in intensity ($\mathcal{O}(10)~$mK in terms of brightness temperature). After that, first stars strongly heat the gas, making it hotter than the 
background radiation and that will manifest itself as an excess in the redshifted line of the neutral hydrogen. The cosmological signal vanishes in the end of the EoR, when the fraction of the neutral hydrogen greatly decreases. 
Thus, evolution of the radiation sources during Cosmic Dawn and Epoch of Reionization would leave a distinctive imprint on the spectrum of the neutral hydrogen line in the approximately 30-200 MHz frequency range. This feature is global so it is possible to search for it using even a single dipole antenna, e.g. \cite{Shaver1999}. There are several experiments underway aimed at the search of the signal from these early epochs:  Experiment to Detect the Global EoR Step (EDGES) \cite{Bowman2008}, Shaped Antenna measurement of the background RAdio Spectrum (SARAS) \cite{Singh2017}, SCI-HI experiment (Sonda Cosmol\'ogica de las Islas para la Detecci\'on de Hidr\'ogeno Neutro)\cite{Voytek2014}, Broadband Instrument for Global HydrOgen ReioNisation Signal (BIGHORNS) \cite{Sokolowski2015}. More ambitious projects are underway as well: Large-Aperture Experiment to Detect the Dark Age (LEDA) \cite{Price2018},  where the monopole signal is searched for using 512-element (simple dipole) full correlation imaging array, Precision Array for Probing the Epoch of Reionization (PAPER) \cite{Parsons2010} and its extension   Hydrogen Epoch of Reionization Array (HERA)~\cite{DeBoer2017}. Observations with these instruments would hopefully reveal higher multipole components of the signal and subsequently make possible to investigate the underlying power spectrum of density fluctuations.

The sought signal is very weak and this pose the major problem for these experiments. The expected level of the signal does not exceed several tens mK when the level of foreground, mostly from the Galactic synchrotron radiation, has an amplitude of several thousand K \cite{deOliveiraCosta2008}. Obviously, the signal extraction is very difficult and can only be done exploiting difference in spectra of the signal and the foreground: while the latter one is smooth and monotone, the shape of the former one is highly non-trivial. It was shown that subtraction of polynomials of sufficiently high orders allows to extract the signal \cite{Pritchard2010}. Another serious obstacle for the searches is artificial radio frequency inference which is rather strong at these frequencies. Some experiments try to ameliorate this problem moving to less populated areas like  Western Australia, other, such as LOFAR based in the Netherlands,  rely on advanced signal processing algorithms. In any case, the importance of precise  calibration can be easily seen, because only then the  estimation of the foreground at the necessary  $10^{-4}-10^{-5}$ accuracy level becomes possible.

Recently, the EDGES experiment announced discovery of a new spectral feature -- an absorption profile centered at 78 MHz and 18 MHz  wide \cite{Bowman2018}. The shape of the profile can be described by certain theoretical models but the estimated amplitude was 0.5 K which is two times higher than the expected value.  This discrepancy could not be explained in the existing  models with 'standard' astrophysics. Thus, almost immediately, a large number of extended models, featuring e.g. Dark Matter with non-trivial properties which help it to cool the gas, was suggested \cite{Barkana2018}.

In this work we approach to the problem from the side of air-shower detection with radio antennas~\cite{Schroder:2016hrv}.
Modern radio detectors are equipped with antennas and electronics operating in the frequency band of 30-80~MHz, which brings the potential of application of air-shower detectors to the problems of EoR signal detection.
Particularly, the scientific program of future GRAND includes these topics~\cite{Alvarez-Muniz:2018bhp}.
The main challenges in these case are the following:
\begin{itemize}
\item Data acquisition in air-shower detectors differs from one using in radio telescopes.
Instead of the integration of the signal, air-shower detectors record the entire uncorrelated traces from the different antennas.
On the one hand it decreases the exposure of the detector, on the other it can improve the quality of the data, since the data contain more information.
\item The systematic uncertainties are crucial for the detection of EoR signal, meanwhile for the air-shower detection the uncertainties in order of 10\% are sufficient.
\end{itemize}
To answer the question of feasibility of the application of air-shower detectors to the detection of EoR signal, we have deployed an engineering array Tunka-21cm aimed at the study of the control of systematics for air-showers detector and tuning data analysis for its redundant data acquisition.

\section{Tunka-21cm setup}
\begin{figure}[t!]
\includegraphics[width=1.0\linewidth]{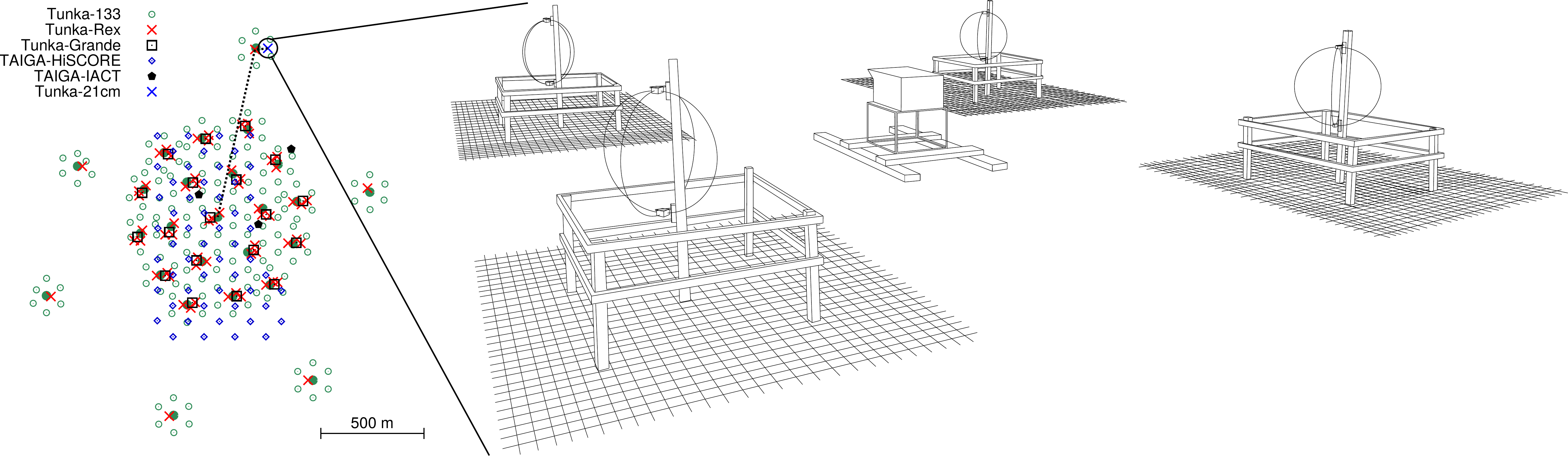}
\caption{\textit{Left:} Tunka-21cm location within TAIGA observatory (depicted by a single marker due to scale of the map), the dashed lines indicate optical fiber connection with the central DAQ.
\textit{Right:} Sketch of Tunka-21cm array depicting cluster center, antenna stations and 4$\times$4~m screens.}
\label{fig:cluster}
\end{figure}

The array is co-located with the Tunka Advanced Instrument for cosmic rays and Gamma Astronomy (TAIGA)~\cite{Budnev:2017fyg,Kostunin:2019nzy} and is installed in northern part of facility (see Fig.~\ref{fig:cluster}).
The array consists of four antenna stations located in the square grid $10\times10$~m, each antenna station is screened from below by a $4\times4$~m grounded metal net.
All antenna stations are equipped with the same electronics and plugged to the ADC in the cluster center.
Digitized data is transmitted to the central DAQ of TAIGA via optical fibers.

\subsection{Antennas, electronics and systematic uncertainties}
\begin{figure}[t!]
\centering
\includegraphics[height=0.36\linewidth]{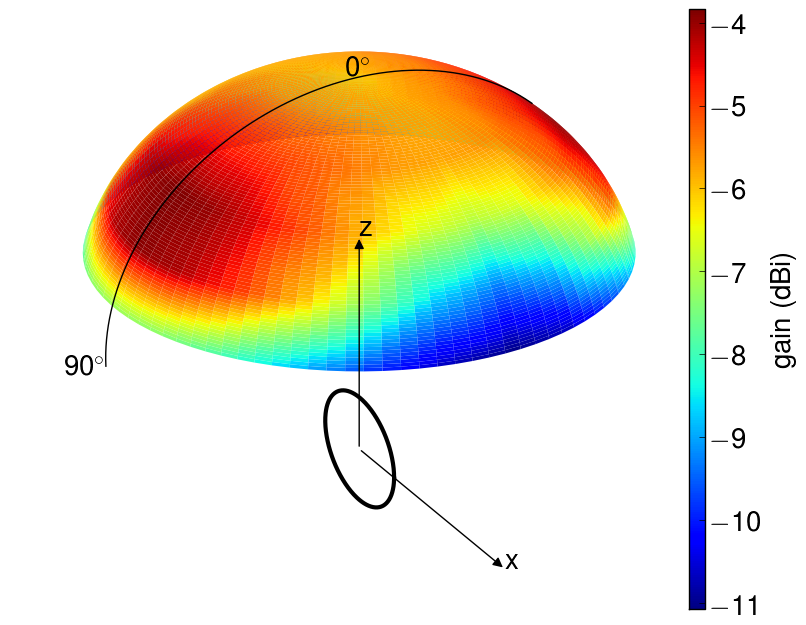}~~~~~
\includegraphics[height=0.36\linewidth]{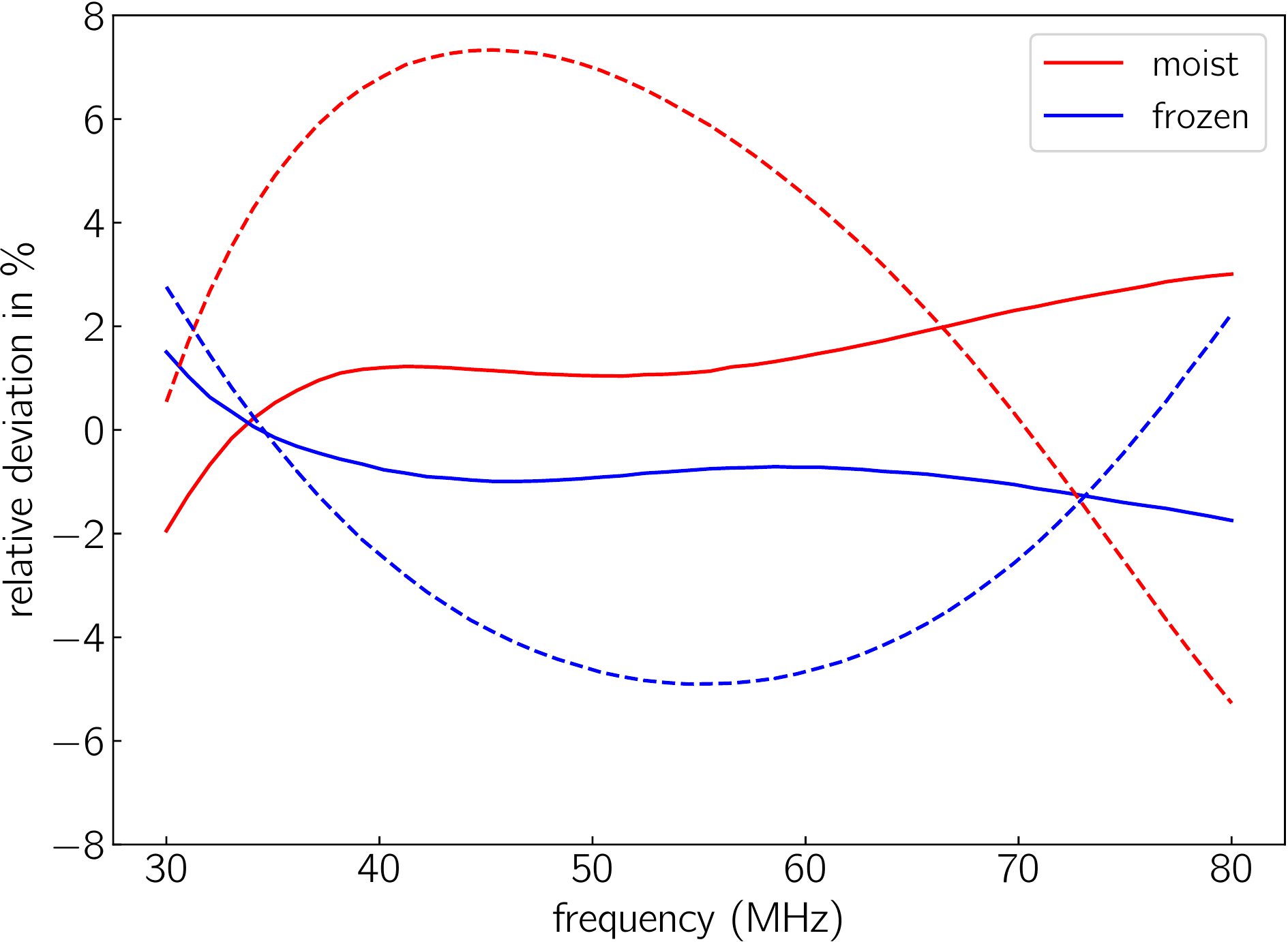}
\caption{\textit{Left:} Exemplary gain pattern of SALLA at 50~MHz.
\textit{Right:} The influence of the ground on the SALLA gain for the Tunka-Rex configuration (no screen, dashed lines) and for the Tunka-21cm configuration (screen $4\times4$~m, solid lines). One can see that the screening significantly decreases the systematic uncertainty from the ground.
}
\label{fig:gain}
\end{figure}

Antenna station design is taken from the Tunka-Rex~\cite{Bezyazeekov:2015rpa} detector.
Single station consists of two perpendicular short aperiodic loaded loop antennas (SALLA)~\cite{KroemerSALLAIcrc2009, Abreu:2012pi}.
The signals from the antennas are pre-amplified with Low Noise Amplifier (LNA) and are transmitted via 15~m coaxial cables to the analog filter-amplifier, which cuts the frequency band to 30-80~MHz.
The signal circuit (LNAs and filter-amplifiers) was calibrated under laboratory conditions.
The antenna pattern and phase response were calculated with the simulation code NEC2~\cite{nec2}.
In Fig.~\ref{fig:gain} one can see the gain pattern of SALLA and influence the screened and unscreened ground on it.
By installation of the screen the systematic uncertainty from the ground is decreased to the maximum of 3\% (if the ground is unknown, however, the approximate ground conditions can be reconstructed from the meteorological data).
During the calculation of the antenna pattern we have figured out the additional numerical uncertainty in order of few percents can be introduces by the choosing of different number of segment for the input, and for the moment we are investigating this.

The cross-check with in-field calibration will be performed with external reference source VSQ 1000 by Schaffner Electrotest GmbH (now Teseq).
This source was also used for the calibration of experiments LOPES, LOFAR and Tunka-Rex.
It consist of biconical antenna DPA 4000 and signal generator RSG 1000.
Source has radiation pattern close to cosine and broadband spectrum (from 30 MHz up to 1 GHz) with narrow peaks at each MHz.
For the calibration we will attach the source to the carbon mast and take the measurements at different positions corresponding to different distances from the antenna and zenith and azimuth angles (as shown in Table~\ref{positions}).

From the experience of Tunka-Rex we can set the upper limits for the following uncertainties:
\begin{itemize}
\item Antenna production and alignment can bring up to 2\%. This value was average over the entire Tunka-Rex array during its live cycle, however for the Tunka-21cm the status of mechanics can be controlled for each run, so we can decrease this uncertainty significantly.
\item Environmental temperature can bring up to 4\%.
With climate-control at the cluster center we can fix the temperatures for the ADC and filter-amplifiers, however the gain of the LNA is affected by the outside temperature.
The influence of the temperature on the gain is measured and known and can be taking into account to cancel this uncertainty.
\item The influence of crosstalk between channels is measured in order of 2\%, this uncertainty is by design and most likely cannot be decreased without significant improvement of the electronics. 
To protect signal chain against external background we developed and installed additional screens on the ADC and filter-amplifiers.
\end{itemize}
Let us not, that in the present work we do not consider the uncertainties, which are not directly related to the hardware of Tunka-21cm, e.g. atmosphere and astronomical sources.

\begin{table}
	\begin{center}
		\resizebox{\textwidth}{!}{
		\begin{tabular}{| c | c | c | c | c | c | c | c | c | c | c | c | c | c | c | c |}
			\hline
				Pos. 	&X (m) 	&Y (m) 	&Z (m) 	&$\theta_1$ 	&$\theta_2$ 	&$\theta_3$ 	&$\theta_4$ 	&$\phi_1$ 	&$\phi_2$ 	&$\phi_3$ 	&$\phi_4$ 	&$D_1$(m) 	&$D_2$ (m) 	&$D_3$ (m) 	&$D_4$ (m) \\ \hline
				a 	&0.00 	&0.00 	&8.70 	&45.11 	&45.11 	&45.11 	&45.11 	&135.00 	&225.00 	&315.00 	&45.00 	&10.02 	&10.02 	&10.02 	&10.02 \\ \hline
				b 	&12.07 	&0.00 	&6.60 	&15.70 	&30.00 	&30.00 	&15.70 	&106.33 	&125.27 	&54.73 	&73.67 	&18.48 	&10.00 	&10.00 	&18.48 \\ \hline
				c 	&8.54 	&8.54 	&10.26 	&31.80 	&60.00 	&31.80 	&24.30 	&75.35 	&45.0 	&14.65 	&45.00 	&16.46 	&10.00 	&16.46 	&21.015 \\ 
			\hline
		\end{tabular}}
	\end{center}
	\caption{Coordinates of the reference source and corresponding radiation angles and distances for all Tunka-21cm antennas.}
	\label{positions}
\end{table}	

\subsection{Data acquisition}

\begin{figure}
\includegraphics[width=1.0\linewidth]{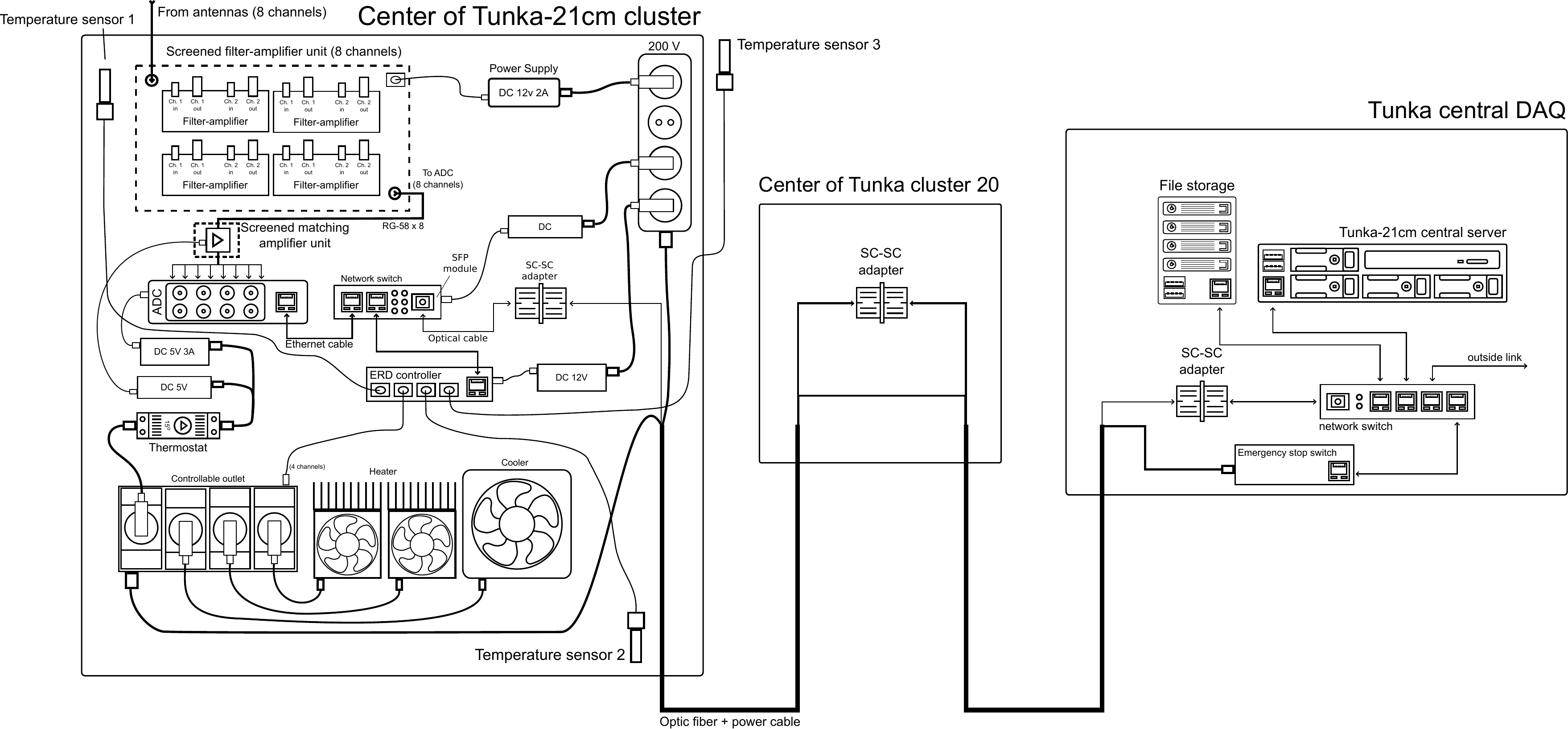}
\caption{The sketch of the Tunka-21cm data acquisition.
One can see, that the chain consists of three nodes: cluster center of Tunka-21cm array with ADC and filter-amplifiers, then link to the Tunka cluster~20 and then existing optic fiber connection between Tunka cluster 20 and central DAQ of TAIGA facility with Tunka-21cm central server and data storage.
This structure was chosen in order to use the existing TAIGA infrastructure with maximum efficiency.
}
\label{fig:tunka21daq}
\end{figure}

For the data acquisition we use ADC board recently developed by SINP MSU.
The board digitizes radio with the sampling rate of 200 MS/s and dynamic range of 12 bits same as for Tunka-Rex, however the length of the traces is about 56~$\mu$s (11$\times$1024 counts), what gives a spectral resolution of about 20~kHz, which is almost an order of magnitude better than for Tunka-Rex.
The order is equipped with the modern Ethernet interface allowing remote control and having a bandwidth of about 30~Mbit/s, what converts to a maximal trigger rate of about 15~Hz and data flow of about 10 GiB/h.
This way, our initial DAQ conditions are similar to the modern and future self-trigger radio detectors.
In this work we do not provide very detailed description of the infrastructure and networking of Tunka-21cm, the sketch of it can bee seen in Fig.~\ref{fig:tunka21daq}

\section{Data analysis}
As it was discussed above, the air-shower detectors acquire redundant data, which allows one to test different modes of the operation and select one more suitable for the detection of EoR signal.
In the frame of Tunka-21cm project we will try the following strategies: single-antenna mode (EDGES-like) will be used as a reference for the cross-check of future improvements; uncorrelated array will lower the threshold in two-three times (as square root of number of antennas) and phased array with taking into account the map of the background will operate similar to the radio telescopes.
With the redundancy of the data and digital beam-forming it is possible to test these modes in parallel and perform direct cross-check of the performance.

The identification of RFI and tagging noised traces are of the great importance for the EoR signal detection.
To solve this problem we will use classical statistical methods and existing open source software for RFI identification as well as modern methods like neural networks, which can be effectively used for the identification of the features of background as it was shown for Tunka-Rex~\cite{Bezyazeekov:2019jbe}.

To compare the performance of the improved air-shower array with screened antennas and controlled systematic uncertainty we are going to compare analysis of the data from Tunka-21cm array with published data from Tunka-Rex.

\subsection{Published Tunka-Rex data}
The Tunka-Rex data is being published in the frame of Tunka-Rex Virtual Observatory (TRVO)~\cite{Bezyazeekov:2019onw}, and we have received access to the pilot version of the client to the TRVO database.
The full database in its final stage will contain all Tunka-Rex traces recorded during seven years of operation~\cite{Kostunin_ICRC2019} and will have a size of order of TiB.
Such large amount of data is an ideal testbed for the benchmarking of data analysis methods for Tunka-21cm.
We have got an access to the pilot version of TRVO and will reanalyze air-shower data with methods developed for Tunka-21cm.

\pagebreak

\section{Conclusion}
Tunka-21cm project is aimed at the proof-of-feasibility of application air-showers detectors operating in 30-80~MHz band to the detection of the EoR signal.
This implies improved control of systematic uncertainties and adaption of the astronomy methods of analysis to the data acquired by air-shower detector.
Having Tunka-Rex detector, which have operated for seven years, we have deployed an improved engineering array based on the existing hardware.
As a result we will directly evaluate the gained performance and accuracy, and can conclude the efficiency of improvements and the feasibility of them.
The obtained results can be useful for the digital arrays operating in 30-80~MHz frequency range aimed either at cosmic-ray detection or at detection of EoR signal.

\section*{Acknowledgements}
This work has been supported 
by the Russian Foundation for Basic Research (grants No. 18-32-00460 and No. 18-32-20220).
We thank colleagues from the TAIGA and Tunka-Rex collaborations for their support during development and deployment of Tunka-21cm array.

\bibliographystyle{ieeetr}
\bibliography{references}


\end{document}